\documentclass[preprints,article,accept,moreauthors,latex]{Definitions/mdpi}
\firstpage{1} 
\pubvolume{xx}
\issuenum{1}
\articlenumber{5}
\pubyear{2019}
\copyrightyear{2019}
\history{Received: date; Accepted: date; Published: date}
\Title{
  Importance of vdW and long-range exchange interactions to
  DFT-predicted docking energies between plumbagin and cyclodextrins
}

\Author{
  Tom Ichibha $^{1,}$*, % ichibha@icloud.com
  Ornin Srihakulung$^{1,2,3}$, % mwkons1501@icloud.com
  Guo Chao$^{1}$, % mwkguc1704@icloud.com
  Adie Tri Hanindriyo$^{4}$, % adietri@icloud.com
  Luckhana Lawtrakul$^{2}$, % luckhana@siit.tu.ac.th
  Kenta Hongo$^{5,6,7,8}$, % kenta\_hongo@mac.com
  and Ryo Maezono$^{1,8}$, % rmaezono@mac.com
}
\AuthorNames{
  Tom Ichibha,
  Ornin Srihakulung,
  Guo Chao,
  Adie Tri Hanindriyo,
  Luckhana Lawtrakul,
  Kenta Hongo,
  and Ryo Maezono
}

\address{
  $^{1}$ \quad School of Information Science, JAIST, Asahidai 1-1, Nomi,
  Ishikawa, 923-1292, Japan; mwkons1501@icloud.com (O.S.); 
  mwkguc1704@icloud.com (G.C.); rmaezono@mac.com (R.M.)\\
  $^{2}$ \quad School of Bio-Chemical Engineering and Technology,
  Sirindhorn International Institute of Technology,
  Thammasat University, Thailand; luckhana@siit.tu.ac.th (L.L.)\\
  $^{3}$ \quad National Metal and Materials Technology Center,
  Pathumthani 12120 Thailand\\
  $^{4}$ \quad School of Materials Science, JAIST, Asahidai 1-1, Nomi,
  Ishikawa, 923-1292, Japan; adietri@icloud.com\\
  $^{5}$ \quad
  Research Center for Advanced Computing Infrastructure, JAIST,
  Asahidai 1-1, Nomi, Ishikawa 923-1292, Japan; kenta\_hongo@mac.com\\
  $^{6}$ \quad PRESTO, JST, Kawaguchi, Saitama 3320012, Japan\\
  $^{7}$ \quad Center for Materials Research by Information Integration,
  Research and Services Division of Materials Data and Integrated System,
  National Institute for Materials Science (NIMS), Tsukuba 305-0047, Japan\\
  $^{8}$ \quad Computational Engineering Applications Unit, RIKEN, 
  2-1 Hirosawa, Wako, Saitama 351-0198, Japan
}

\corres{Correspondence: ichibha@icloud.com}

\abstract{
  We calculated the docking energies between plumbagin
  and cyclodextrins, using density functional theory (DFT)
  with several functionals and some semi-empirical methods.
  Our DFT results revealed that GD3 dispersion force
  correction significantly improves the reliability of prediction.
  Also sufficient amount of long-range exchange is important
  to make it reliable further, agreeing with the previous work
  on argon dimer.
  In the semi-empirical methods, PM6 and PM7
  qualitatively reproduce the stabilization
  by docking , yet under- and over-estimating
  the docking energies by $\sim$10~kcal/mol,
  respectively.
}


\begin{document}
%%%%%%%%%%%%%%%%%%%%%%%%%%%%%%%%%%%%%%%%%%%%%%%%%%%%%
\section{Introduction}
\label{sec.intro}
%%%%%%%%%%%%%%%%%%%%%%%%%%%%%%%%%%%%%%%%%%%%%%%%%%%%%
Biopharmaceuticals are manufactured, extracted, and
semi-synthesized from biological sources.
Compared to chemically synthesized pharmaceuticals,
they have high potency at low dose and possess
distinctive medical properties.\cite{2014MIT, 2006GEO}
On the other hand, they often lack physical and
chemical stability, which poses problems for
long-lasting storage, oral ingestion, and so on.
One of the most promising ways to solve the problem
is host-guest docking technology.\cite{2006GEO,2014MIT}
The biopharmaceutical molecules (guest) are combined with the carrier 
molecules (host) and are stabilized both physically and chemically.
It also allows one to control where and how fast
the biopharmaceutical is released from the carrier
and absorbed in human/animal body.\cite{2008VYA}

%反面、「薬剤を安定的に保存するのが難しい/コストを要する」、
%「胃酸に耐えないため経口にて投与できず注射での投与が必要となる」
%といった扱いづらさがある\cite{2014MIT, 2006GEO}。
%バイオ薬品は生体の作用を用いて製造、抽出、半合成される医薬品である。
%化学合成医薬品と較べて、少ない投与量で高い薬効を発揮する、
%その薬固有のユニークな薬効を有するといった傾向がある\cite{2014MIT, 2006GEO}。
%此等の問題に対する有望な方策としてドッキング技術が挙げられる
%\cite{2006GEO, 2014MIT}。薬剤物質を、キャリアとなる分子と
%結合させることで、物理的・化学的に安定化させる。
%加えて、薬剤放出のレートやタイミングを管理することも可能となる

\vspace{2mm}
The binding energy of the docking is of great
relevance to the stability and the release rate/timing
of the biopharmaceuticals.
Density functional theory is expected to be able to
provide theoretical prediction with high reliability beyond
those of semi-empirical methods.\cite{2016CHR}
%安定性や薬剤放出レート/タイミングの予見には、密度汎関数法(DFT)を
%用いたエネルギー評価が期待されている。
Yet, even for DFT, it is difficult to make a reliable
energy estimation since multiple non-covalent forces
such as hydrogen bond or dispersion force are complexly
intertwined to realize the docking interaction.\cite{2016YE}
%しかしながら、ドッキングは
%ファンデルワールス力や水素結合といったnon-covalent力が複雑に
%絡み合って実現されるものであり\cite{2016YE}、このような状況を
%密度汎関数を用いて正確に記述することは非常に難しい。
In other words, special treatments for the non-covalent
forces are needed.
One example is long-range correction,\cite{2001IIK,2004YAN}
which enhances the proportion of the exact exchange term
for long-range interactions, which can improve the description
of van der Waals forces.\cite{2012TSU}
%First, 長距離補正\cite{2001IIK,2004YAN}では、長距離の
%相互作用における厳密交換の割合をエンハンスすることでファンデル
%ワールス力の記述性を向上させることを目指している
Another example is using a family of Minnesota functionals,
whose parameter training-set includes weak-force interaction
systems and has an improved accuracy to describe such systems.\cite{2007ZHA}
%Second, 一連のミネソタ汎関数では、較正パラメータは「弱い結合力が支配的な系」
%について調教されており、長距離のnon-covalent力の記述精度を大幅に
%向上させることに成功している\cite{2007ZHA}。
However, they cannot describe the asymptotic decline of van der Waals
forces in proportion to $R^{-6}$~($R$: inter-atomic distance), since
they do not explicitly contain dispersion interactions by its construction
and also their internal parameters are optimized only around equilibrium
geometries.\cite{2017HON}
This decline can be described by Grimmes's dispersion
correction (GD3),\cite{2010GRI}  which adds an empirical 
function akin to the Lennard-Jones potentials\cite{2011GRI} and
systematically improves the description of van der Waals systems.
\cite{2010GRI,2011BUR}
%これに対し、Grimme's dispersion correction (GD3) 
%Third, Grimme's分散力(GD3)補正では、経験的ポテンシャルを用いて、
%分散力をはじめとするnon-covalent力の補正を行っており\cite{2011GRI}、
%当該力が支配的な系でのエネルギー予見精度を系統的かつ劇的に改善する
%ことが報告されている\cite{2011GRI}{\color{red}[文献を足す]}。

\vspace{2mm}
For several systems, especially for host-guest docking of biopharmaceutical
compounds, system size is often an insurmountable obstacle for conventional
Kohn-Sham(KS)-DFT applications.
On the other hand, the recent progress of DFT out of the conventional KS-DFT
has shown promise to realize protein-scale DFT calculation.
\cite{2018ROM,2015MOH}
Therefore, it is a significantly meaningful task to study what exchange-correlation
functional works for the host-guest docking in order to prepare for
similar large scale calculations.
%また薬剤分子を対象とした場合には、しばしば大きすぎるという
%ことが問題となるだろう。一方で、近年の試みでは、DFTの
%スケーリング性能を上げるようにしているので、この問題は解決
%されることが期待される。そうした場合には、どの汎関数で良く
%記述できるかというのが、非常に重要になってくる。

\vspace{2mm}
In this work, we tested eight functionals
listed in Table \ref{tab.functionals}, targeting
the binding energy between plumbagin and cyclodextrins. 
This system is a representative example of the host-guest docking
of biopharmaceutical compounds, providing a reasonably-sized
system for the conventional KS-DFT calculation.
We concluded the co-existence of vdW correction
and sufficient amount of long-range exchange is
essential for having a quantitatively reliable prediction.
In addition to DFT, we applied semi-empirical
methods PM3, PM6, and PM7,\cite{2016CHR} and found
PM6 and PM7 reproduce the stabilization of docking
qualitatively.
Nevertheless, they under- and over-estimate the docking
energy by $\sim$10~kcal/mol, compared to the best
DFT predictions.

%%%%%%%%%%%%%%%%%%%%%%%%%%%%%%%
\begin{table}
  \begin{center}
    \caption{\label{tab.functionals}%[tab.functionals]
      List of exchange-correlation functionals we tested.
      We examined the reliability of common functionals,
      B3LYP, M06L, and M06-2X, and their relatives with
      GD3 and/or CAM corrections for predicting
      the docking energies between plumbagin and cyclodextrins.
    }
    \begin{tabular}{c|c|c|c}
      Plain  & GD3        & CAM       & CAM+GD3       \\
      \hline
      B3LYP  & CAM-B3LYP  & B3LYP-GD3 & CAM-B3LYP-GD3 \\
      M06L   & M06L-GD3   & --        & --            \\
      M06-2X & M06-2X-GD3 & --        & --            \\
    \end{tabular}
  \end{center}
\end{table}
%%%%%%%%%%%%%%%%%%%%%%%%%%%%%%%
%%%%%%%%%%%%%%%%%%%%%%%%%%%%%%%
\section{System}
\label{sec.system}%[sec.system]
%%%%%%%%%%%%%%%%%%%%%%%%%%%%%%%
Plumbagin is an organic molecule including two benzene rings,\cite{1999OOM}
which is reported to be effective against prostate cancer.\cite{2008AZI,2017ABE}
However, this molecule cannot exist for a long time under normal atmospheric conditions
as oxidation and degradation can cause losses of up to 63.8\% in a single month.
\cite{2010SUT} Docking plumbagin molecules within cyclodextrin has been considered
as an effective method to extend its short shelf-life.
%プルンバギンは2つのベンゼン環を含む有機分子であり\cite{1999OOM}、
%プルンバゴ類の薬草から見つかった薬剤分子で前立腺がんへの薬効が期待されている
%\cite{2008AZI,2017ABE}。
%しかしながら、プルンバギンには、室温下で長期間安定に存在することが
%できないという弱点があり、昇華や酸化等によるdegradationによって
%一ヶ月で63.2\%も失われてしまうが\cite{2010SUT}、シクロデキストリン類
%とのドッキングにより大幅に抑制できることが知られている。

\begin{figure}[htbp]
  \centering
  \includegraphics[width=1.0\hsize]{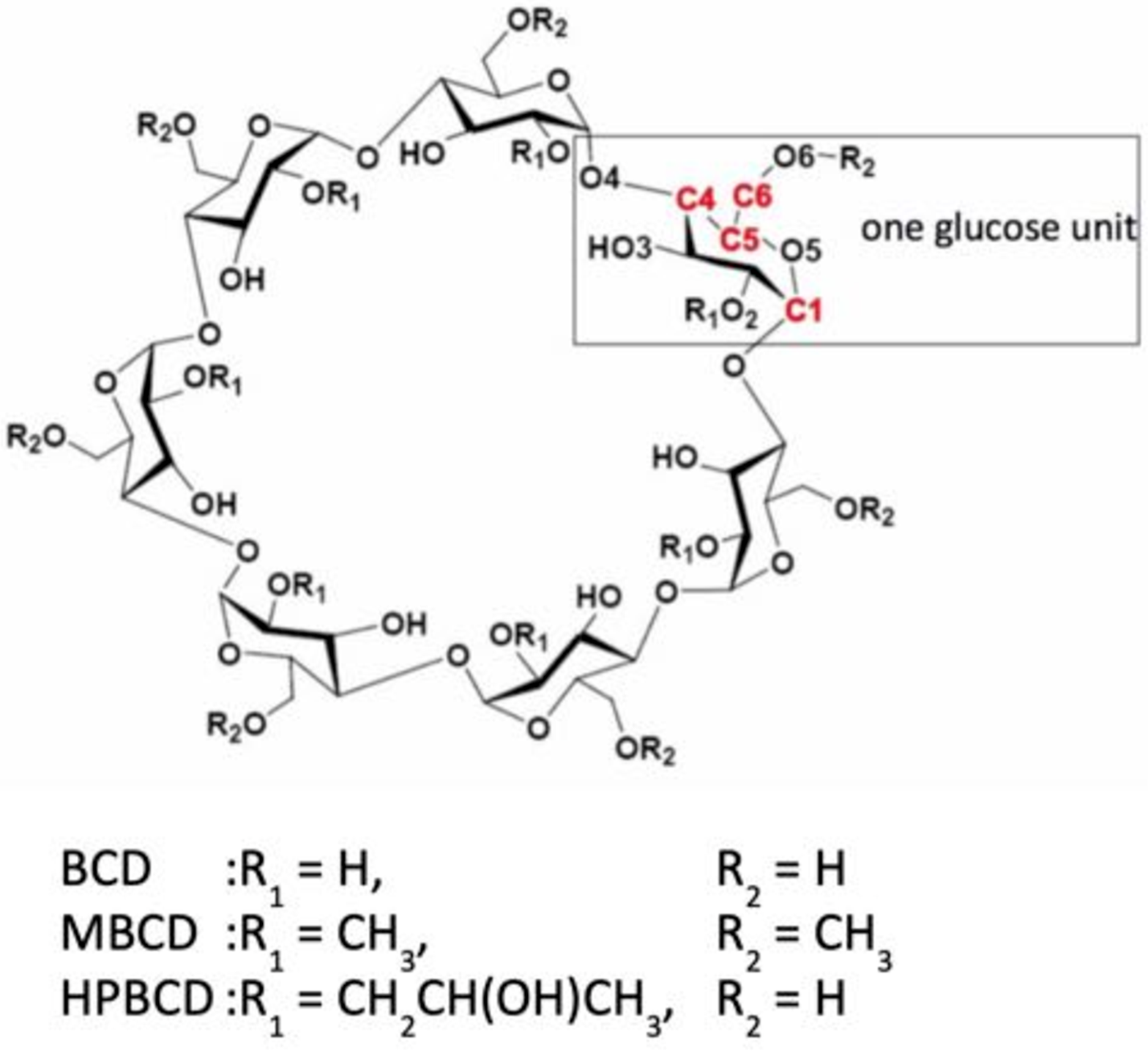}
  \caption{
    \label{fig.structure}%[fig.structure]
    The molecular structure of BCDs.
    The ring consists of glucose units.
    There are a variety of BCDs according to
    the functional groups located at R$_1$ and R$_2$.
    We selected BCD, MBCD, and HPBCD,
    shown in this figure, and calculated
    the binding energy when docked with plumbagin.    
  }  
\end{figure}

\vspace{2mm}
Cyclodextrin (CD) is a circular molecule of glucose units as shown in
Figure~\ref{fig.structure}. It is broadly used as a carrier of
pharmaceuticals due to the following merits (other than stabilization):
\cite{2008VYA}
%シクロデキストリン類は図\ref{fig.structure}に示すような環状分子である。
%当該キャリアは、分子を安定させる以外にも以下のような利点のために幅広く
%応用されている:\cite{2008VYA}
\begin{itemize}
\item
  The ring size is adjustable to the size of guest molecule
  by changing the number of glucose units $n$ ($\ge$6).
  For $n$ equaling 6, 7, and 8, it is called
  $\alpha$-, $\beta$-, and $\gamma$-CD, respectively.
\item
  Docking with CD improves drug solubility or dissolution,
  which is essential for drugs with poor water solubility.
\item
  The release rate/timing is controllable by assigning
  different functional groups at $R_1$ and $R_2$ in Figure \ref{fig.structure}.
\end{itemize}
We selected $\beta$-CD (BCD) as the host molecule, since
it has been experimentally used for docking with plumbagin,\cite{1999OOM}
and we obtained as well the docking energies for plain BCD
and its two variants, Methyl-BCD (MBCD) and Hydroxy Propyl-BCD (HPBCD),
shown in Figure \ref{fig.structure}.

%%%%%%%%%%%%%%%%%%%%%%%%%%%%%%%
%%%%%%%%%%%%%%%%%%%%%%%%%%%%%%%%%%%%%%%%%%%%%%%%%%%%%
\section{Methods}
\label{sec.details}
%%%%%%%%%%%%%%%%%%%%%%%%%%%%%%%%%%%%%%%%%%%%%%%%%%%%%
We obtained the docking structures between plumbagin
and BCDs from docking analysis with Lamarckian algorithm
using AutoDock 4.2.6.\cite{1998MOR}
This algorithm is often used to predict the ligand
arrangements of protein systems.
A set of genes representing the ligand arrangements 
are optimized to get energetically stable structures.
Each of the genes consists of translations, orientations,
and conformations of the ligands. We optimized the
arrangement of plumbagin as the `ligand' of BCDs.
%With the lamarkian algorithm, a few percent of the genes
%are relaxed by force calculations, which significant
%プルンバギンとシクロデキストリン類のとのドッキング構造は、
%AutoDock4.2.6に実装されている、ラマキアン遺伝的アルゴリズムに
%基づく手法から与えた\cite{1998MOR}。当該手法は、タンパク質と
%結合するリガンド配置の探索に多用されている。当該アルゴリズムは
%タンパク質のリガンド配置のタンパク質に対する各リガンドの位置(translation)、
%角度(orientation)、ねじれ(conformation)を「遺伝子(genotype)」とし、
%エネルギーの低い構造を目指してgeneをアップデートしていく。本計算では
%シクロデキストリン類をプルンバギンに対するリガンドとして取り扱って
%ドッキング構造の探索を行った。

\vspace{2mm}
The molecular structures of plumbagin and BCDs are
taken from the entries: PVVAQS01, BCDEXD03, BOYFOK04,
and KOYYUS in the Cambridge Structural Database.\cite{2016GRO}
We optimized their structures using semi-empirical PM7
before doing docking analysis.
In the docking analysis, the translation of plumbagin
is discretized on a 50$\times$38$\times$24 grid with
spacings of 0.375~$\AA$. We run 100 iteration for 150
of initial population of genes.
At the end of each iteration, we selected only one
gene with the lowest energy to "survive" to the next iteration.
The energies were calculated with empirical force field,
where electrostatic interaction was given based on
Gasteriger charges.~\cite{1978GAS}
The other input parameters were set to be the default
values of Autodock 4.2.6.
%After 100 iterations,
%the energies of the genes are converged in the range of
%{\color{red}XXX}~kcal/mol.

%\begin{comment}
%\vspace{2mm}
%まず、通常の遺伝的アルゴリズムについて説明する。
%当該アルゴリズムでは、genotypeを一回更新するのに、mapping、selection、
%crossover、mutation、elite selectionの5段階を経る。まず、探索の開始時に、
%genotypeを複数個ランダムに生成する。mappingではgenotypeを対応するリガンド配置
%(phenotype)へと変換し、そのエネルギー的安定性を半経験的手法などから評価する。
%selectionではエネルギーに応じて各genotypeを複製する。この際、エネルギー的に
%安定なものほど多く複製される。cross-overでは任意回数だけランダムに2つgenotypeを
%抽出し、genotypeの一部を互いに交換することで新たに2つのgenotypeを作る。この際、
%オリジナルのgenotypeは消えるため、genotypeの総数は変化しない。mutationでは
%各genotypeの要素に0を中心とするコーシー分布に従う乱数を足す。当該分布は正規分布に
%似ているが、裾野部分が厚いので大きな変位を与える可能性がある。elite-selectionでは
%当該世代のgenotypeの中でエネルギー的に安定なものを任意個数選びだし、
%次のイテレーションの入力とする。
%
%\vspace{2mm}
%ラマキアン遺伝的アルゴリズムでは、構造最適化を利用し、探索効率の向上を図っている:
%通常の遺伝的アルゴリズムでは、mutationがポテンシャル曲面上の極小点にある構造の
%探索を担っており、それには数(several)世代を必要とする。一方、ラマキアン遺伝的
%アルゴリズムでは、構造最適化によりたった一世代で極小点の構造を得られる。具体的な
%アルゴリズムへの変更としては、mapping後に一部の構造を最適化し、それに応じて
%genotypeおよびエネルギー評価値を更新する。
%最適化後に得られたエネルギーでgeno-typeの評価値を置き換える。また、所与の構造に対して、
%geno-typeが一意に決まる場合に限り、geno-typeを最適化された構造に対応するもので
%置き換える。
%\end{comment}

%%%%%%%%%%%%%%%%%%%%%%%%%%%%%%%%%%%%%%%%%%%%%%%%%%%%%
\vspace{2mm}
We performed DFT calculations using Gaussian09/16.
\cite{g09, g16}
We used the 6-31G++($d$,$p$) basis set, since a family of
6-31G basis sets are often used for docking systems.
\cite{2013BAC, 2014BAC, 2016YE, 2017BOU, 2017DEK}
We corrected the basis set superposition error
by the counterpoise method.\cite{1996SIM}
In addition to the common functionals B3LYP, M06-2X,
and M06L, we compared the reliability of CAM-B3LYP
with long-range exchange correction\cite{cam} and
B3LYP-GD3, M06-2X-GD3, M06L-GD3, and CAM-B3LYP-GD3
with Grimme's dispersion correction,\cite{gd3}
as shown in Table \ref{tab.functionals}.
GD3 correction introduces a pair-wise function akin to the
Lennard-Jones potential to the original functional.
Thus, the corrected functional is no longer within
the framework of density functionals.
%It is reported by the measurements on the S22 benchmark
%system set that GD3 correction systematically improves
%the reliability of prediction.\cite{2010GRI,2011BUR}

%密度汎関数理論(DFT)計算には
%計算コードGaussian 09/16\cite{g09, g16}を用いた。
%ドッキング系の計算には6-31Gをベースとした基底系が多用されており
%\cite{2013BAC, 2014BAC, 2016YE, 2017BOU, 2017DEK}、
%本研究では分極関数と分散関数が含まれるバリエーションである、
%6-31G++($d$,$p$)を用いた。基底重なり誤差の補正にはカウンター
%ポイズ法を用いた\cite{1996SIM}。交換相関汎関数には、B3LYPや
%M06-2X\cite{m06_m062x}といった汎用的に用いられるものに加え、
%分散力補正\cite{gd3}を施したB3LYP-GD3/M06-2X-GD3や
%電荷分極の記述性向上を目指して作られたCAM-B3LYP/CAM-B3LYP-GD3
%\cite{cam}ドッキングエネルギー予見を行った。GD3補正では汎関数に
%「原子間距離に依存したpair-wiseポテンシャル」を加えることから
%通常の密度汎関数の枠組みを逸脱している。\cite{2011BUR}
%当該補正はS22セットに対するベンチマークにおいて、予見信頼性を
%どの汎関数に補正を与えた場合でも系統的に向上させることが報告
%されている\cite{2010GRI,2011BUR}。

\vspace{2mm}
Semi-empirical methods are generally used to predict the structures
of large-size systems represented by proteins.\cite{2016CHR}
Its cost and reliability is lower than DFT but higher than
the methods based on practical potentials. This method performs
one-electron integrals only, so the scaling of the calculation
cost is kept from rising above $\mathcal{O}(N)$ ($N$: number of electrons).\cite{2016CHR}
%半経験的手法はタンパク質に代表されるスケールの大きな有機分子系の構造予見に
%主に用いられる手法である\cite{2016CHR}。計算コスト、および、計算精度の
%観点から、第一原理計算と経験的ポテンシャルとの中間的な手法である。具体的には、
%電子軌道間の積分のうち、一電子積分以外を一電子積分の結果に依存する関数で
%置き換えることで、計算コストを電子数$N$に対して$O(N^1)$にまで抑えている\cite{2016CHR}。
We employed three kinds of semi-empirical methods\cite{2016CHR}
and calculated the docking energies using Gaussian09/16.\cite{g09,g16}
These methods are in the orders of PM3, PM6, and PM7:
PM6 improves on PM3 by refining the core-core interaction term and
introducing the $d$-type basis function.\cite{2016CHR}
%First PM3 to PM6, the core-core interaction term is refined
%and also $d$-type basis function is newly introduced.\cite{2016CHR}
PM7 method further improves on this by introducing corrections for
dispersion and hydrogen bonding.\cite{2013STE}
%From PM6 to PM7, the dispersion and hydrogen bonding
%correction is introduced.\cite{2013STE}

%本研究では半経験的手法のPM3、PM6、PM7\cite{2016CHR}
%を計算コードGaussian\cite{g09,g16}を用いて実行し、ドッキングエネルギーを算定した。
%これらの手法はPM3$\to$PM6$\to$PM7で古く、以下、それぞれの差異について説明する。
%まず、PM3$\to$PM6では、基底関数に新たにスレータ型$d$基底関数を加えることで、
%軌道の記述精度を向上させると共に、core-core相互作用表現を見直している。
%\cite{2016CHR}。
%しかしながら、依然として十分な記述性を得られなかったので、PM6$\to$PM7では
%原子格間距離に依存したペアワイズ関数による分散力補正が新たに追加された
%\cite{2013STE}。

%%%%%%%%%%%%%%%%%%%%%%%%%%%%%%%
\begin{figure}[htbp]
  \centering
  \includegraphics[width=1.0\hsize]{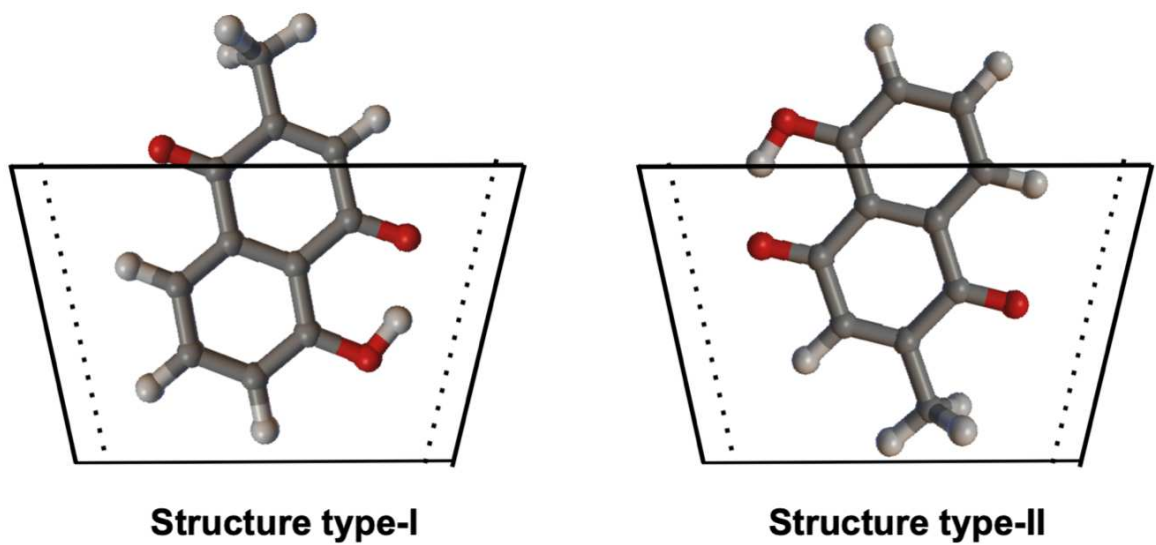}
  \caption{
    \label{fig.up_down}%[fig.up\_down]
    Two types of stable conformations
    found by the docking analysis.
    In type-I(II), the hydroxyl phenolic
    (methyl quinone) group of plumbagin is
    placed around narrow-side of the cavity in BCDs.
  }  
\end{figure}
%%%%%%%%%%%%%%%%%%%%%%%%%%%%%%%%%%%%%%%%%%%%%%%%%%%%%
\begin{figure}[htbp]
  \centering
  \includegraphics[width=1.0\hsize]{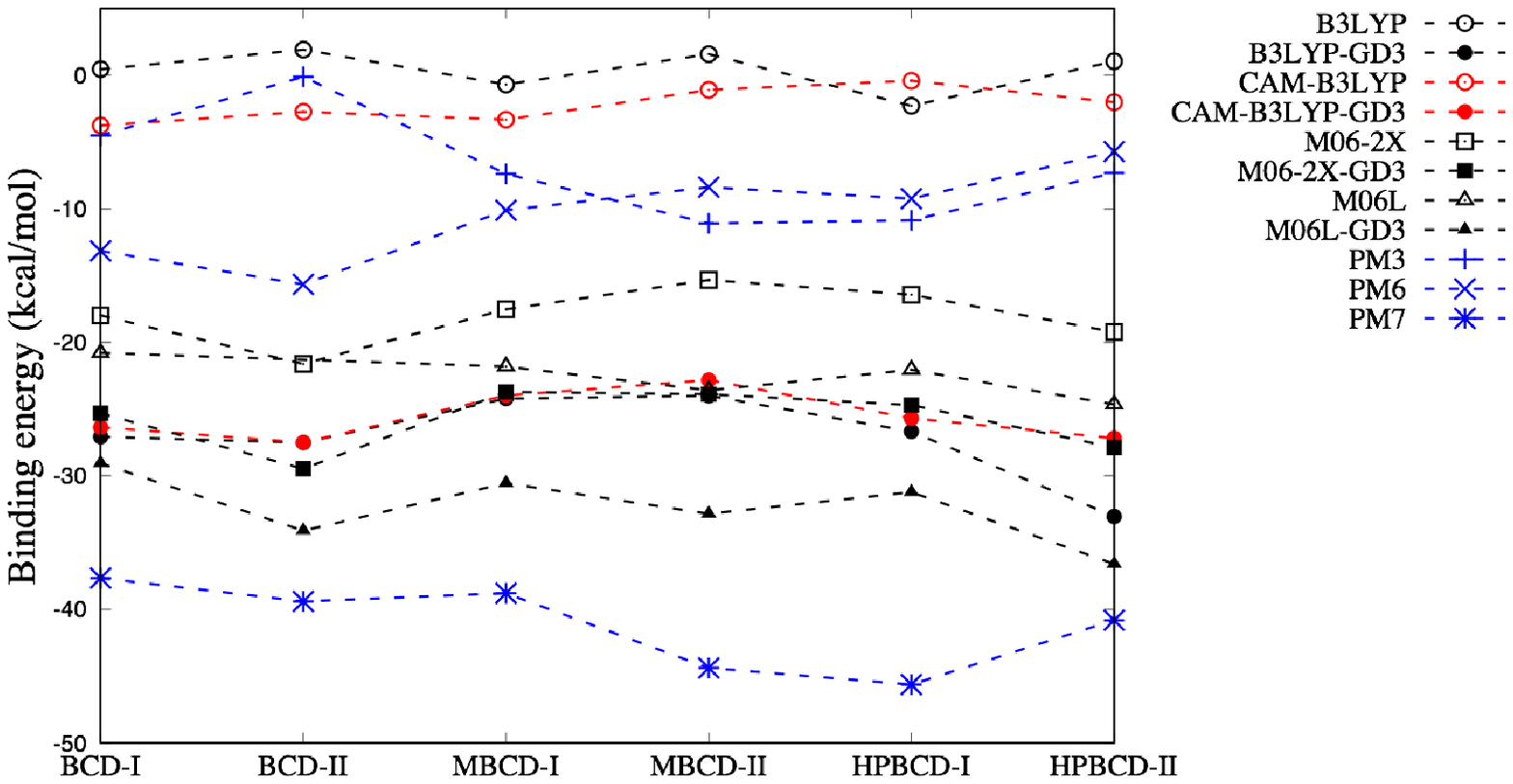}
  (a) All predictions
  \includegraphics[width=1.0\hsize]{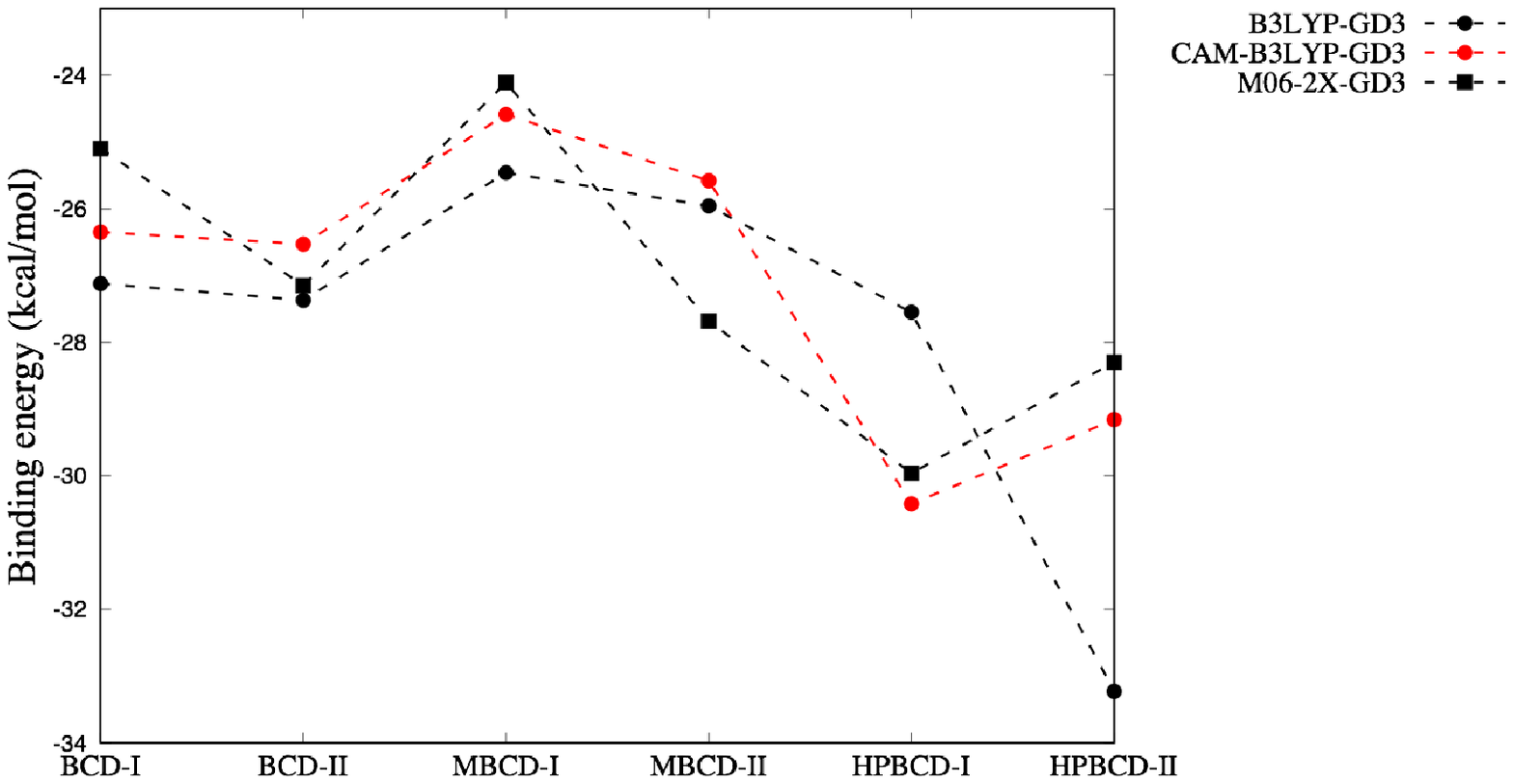}
  (b) GD3 predictions  
  \caption{
    %[fig.energies]
    Comparison of the docking energies predicted by
    DFT with several functionals and the semi-empirical methods.
    Figure (a) shows all of the results and
    figure (b) shows the results obtained from functionals
    with GD3 correction, except M06L-GD3.
    The difference of structure types I and II are
    explained in Figure \ref{fig.up_down}.
    %  
    %    
    %
    %    DFT計算と半経験的手法から得られたドッキングエネルギー予見の比較。
    %    AutoDockから得られた一つのドッキング構造に対し、夫々の手法から
    %    構造最適化を行い、ドッキングエネルギーを算出した。
    %    シクロヘキサン類(BCD, MBCD, PBCD)について、それぞれ構造を
  %    2通り用意しているが、此等の違いについては節\ref{sec.system}で
    %    述べている。
  }
  \label{fig.energies}
\end{figure}
%%%%%%%%%%%%%%%%%%%%%%%%%%%%%%%%%%%%%%%%%%%%%%%%%%%%%
\section{Results and Discussions}
\label{sec.results}%[sec.results]
%%%%%%%%%%%%%%%%%%%%%%%%%%%%%%%%%%%%%%%%%%%%%%%%%%%%%
We found that the structures obtained by the docking analysis
are classified into two types of conformations as shown in
Figure \ref{fig.up_down}. In type-I~(II), the hydroxyl phenolic
(methyl quinone) group of plumbagin is placed around narrow-side
of the cavity in BCDs. Thus, we took both types of conformations
for our benchmark.

\vspace{2mm}
Figure \ref{fig.energies} shows the predictions
of the docking energies between plumbagin and
BCDs for type-I and II by DFT with the functionals
listed in Table \ref{tab.functionals} and the semi-empirical methods. 
First, in comparing the DFT results from functionals without GD3 correction, 
we observe that M06L and M06-2X reproduce the stabilization by docking 
qualitatively, while B3LYP and CAM-B3LYP do not.
This is recognized to be originating from the artifact of (CAM-)B3LYP,
which are optimized just for covalent systems,
resulting in a poor description for non-covalent forces.\cite{2011MAR}
Similar problems have been observed for the stacking of b-DNA\cite{2018KEN}
and the bonding of NiPc dimer.\cite{2011MAR}

%図\ref{fig.energies}では、節\ref{sec.intro}で挙げた
%各種交換相関汎関数をを用いたDFT予見、および、半経験的手法から
%得られた予見結果を比較している。まず、GD3補正無しの汎関数
%(B3LYP, CAM-B3LYP, M06L, M06-2X)でのDFT予見を比較
%すると、(CAM-)B3LYPはドッキングによる安定化を定性的に再現
%出来ていないことが見て取られる。
%同様の失敗はb-DNAスタッキング\cite{2018KEN}やNiPcダイマー
%の結合\cite{2011MAR}でも報告されている。
%(CAM-)B3LYP汎関数のパラメータのトレーニングセットには、
%弱い結合が支配的な系が含まれておらず、それらの記述性が芳しくない
%ことが指摘されており\cite{2011MAR}、本計算でも其れが災い
%したものと考えられる。

\vspace{2mm}
DFT results for functionals with GD3 correction show
that all functionals, with the notable exception of M06L-GD3,
give similar predictions.
It is also most surprising that the predictions 
from CAM-B3LYP-GD3 and M06-2X-GD3 show almost
the same tendencies (see Figure \ref{fig.energies}b).
%It would not be considered that two functionals
%based on different design concepts just accidentally
%shows the coincidence. Rather, it would be natural
%that their predictions are closer to the true docking
%energies so they agrees with each other.
It is unlikely to be a product of coincidence that 
two functionals based on different design concepts
accidentally converge to the same results.
Rather, it seems natural to conclude that
the two functionals predicted very close values
to the true docking energy, and resulting in
the coincidence.
%次に、GD3補正が施された場合の結果を比較すると、M06L-GD3を除き、 
%予見結果が定量的に良く似通っているのが興味深い。とりわけ、
%CAM-B3LYP-GD3とM06-2Xの予見結果は定量的にほぼ一致している。
%構成の異なる2つの汎関数より此のような一致が得られたのは
%偶然とは考えにくく、むしろ
%「それぞれ独立に真の結合エネルギーに近い値を予見している
%  ことから近い予見がえられた」と考えるのが自然であり、
%非常に信頼性の高い結果が得られているものと結論付ける。
Here, the next important question is why these functionals
work better than M06L-GD3 and B3LYP-GD3.
This would be attributed to their having larger
amount of exact exchange used to describe
long-range interactions:
%ここで、B3LYP-GD3、M06L-GD3では、予見信頼性が劣って
%しまったのかが疑問となるが、これは長距離交換の不足による
%ものと考えられる:
Kamiya {\it et al.} established in their benchmark calculations
on the bonding of argon dimer that the balanced
evaluation of van der Waals correlation and long-range
exchange is significant to reliably describe the van der
Waals interaction.\cite{2002KAM}
They observed as well that argon dimer is over-bound
when the long-range exchange is lacking.\cite{2002KAM}
It is in agreement with our results, which shows M06L-GD3
without exact exchange predicts much higher docking
energy than the other GD3 functionals.
%Kamiyaらは、アルゴン二量体の結合
%エネルギーを対象にベンチマーク計算を行い、ファンデル
%ワールス力の記述にはファンデルワールス相関相互作用と
%長距離交換効果のバランスが重要であると結論付けた\cite{2002KAM}。
%特に、長距離交換相互作用を欠く場合にはオーバーバインド
%されることを予見しており、
%厳密交換を含まないM06L-GD3で
%結合が深くなったこととも一致する。

\vspace{2mm}
Finally, we discuss here the results given by PM3, PM6, and PM7
semi-empirical methods.
First, looking at PM3 and PM6, only PM6 reproduces the stabilization
by docking with BCDs. This would be attributed to the
refinement of the core-core interaction in PM6 compared to PM3.
PM7 also reproduces the stabilization for all six patterns.
However, we observe that the binding energies
predicted by PM6 (PM7) are smaller (larger)
by $\sim$10~kcal/mol than those produced from the more rigorous
{\it ab initio} methods (DFT with CAM-B3LYP-GD3 and M06-2X-GD3).
The failures of PM6 and PM7 are due to not explicitly
containing the dispersion force correction and lacking
exact exchange, respectively.
%次に、PM6$\to$PM7ではファンデル
%ワールス補正のためにドッキングに伴う安定化を定性的に
%再現することに成功している。しかしながら、「参照標準」
%と較べてドッキングエネルギーを大幅に過大評価している。
%M06L(-GD3)汎関数と同じく、非局所な交換効果が
%一切考慮されていないことが、ドッキング・エネルギーの
%過大評価に繋がったものと考えられる。

%%%%%%%%%%%%%%%%%%%%%%%%%%%%%%%
\section{Conclusion}
\label{sec.conc}%[sec.conc]
%%%%%%%%%%%%%%%%%%%%%%%%%%%%%%%
We have investigated various types of functionals which
give a reliable estimation of the binding energy of docking
between plumbagin and BCD, MBCD, and HPBCD,
which are representative host-guest docking systems
in the biopharmaceutical field.
Comparing the predictions of non-GD3 functionals,
we find functionals M06L and M06-2X reproduce the stabilization
by host-guest docking while B3LYP and CAM-B3LYP do not.
This is due to B3LYP and CAM-B3LYP having been optimized
just for covalent systems.
Among functionals with GD3 corrections,
%B3LYP-GD3, CAM-B3LYP-GD3, M06-2X-GD3, and M06L-GD3,
we concluded that CAM-B3LYP-GD3 and M06-2X-GD3 predict 
the binding energies very reliably, since they give
surprisingly similar predictions, which cannot be
considered to be just accidental.
This would be attributed to both functionals
possess sufficient amount of long-range exchange
to properly describe non-covalent forces.
Lastly, from the semi-empirical methods, PM6 and PM7
reproduce the stabilization by host-guest docking.
However, each method under- and over-estimates
the binding energy by $\sim$10kcal/mol, respectively.
%我々は、ドッキング系のモデルケースとして、プルンバギンと
%シクロデキストリン類との結合エネルギーを、様々な汎関数を
%用いたDFT計算から予見し、どの汎関数がドッキング系で信頼性
%の高い予見を与えるか調べた。
%まず、グリムの分散力補正の
%施されていない、B3LYP、CAM-B3LYP、M06L、M06-2Xの比較では、
%前者2つは弱い結合力が支配的な系に対して調教されていないため、
%ドッキングによる安定化を再現することが出来なかった。
%次に、
%此等に分散力補正を施した場合の比較では、CAM-B3LYP-GD3と
%M06-2X-GD3が非常に定量的信頼性の高い予見を与えた。これは、
%長距離交換が十分な割合で含まれているためである。

%最後に半経験的手法PM3,PM6,PM7の比較では、PM7のみ
%ドッキングによる安定化を定性的に再現したものの、
%結合エネルギーを10kcal/molも過大評価している。

%%%%%%%%%%%%%%%%%%%%%%%%%%%%%%%
\section{Acknowledgments}
%%%%%%%%%%%%%%%%%%%%%%%%%%%%%%%
The computation in this work has been performed 
using the facilities of the Research Center for Advanced Computing Infrastructure (RCACI) at JAIST.
T.I. is grateful for financial suport from Grant-in-Aid for JSPS Research Fellow (18J12653).
O.S. is grateful for financially supported by SIIT-JAIST Dual degree
scholarship from Thailand's National Electronics and Computer
Technology Center (NECTEC), Sirindhorn International Institute
of Technology (SIIT) and Japan Advanced Institute of Science
and Technology (JAIST).
K.H. is grateful for financial support from a KAKENHI grant (JP17K17762),
a Grant-in-Aid for Scientific Research on Innovative Areas ``Mixed Anion'' project (JP16H06439) from MEXT, 
PRESTO (JPMJPR16NA) and the Materials research by Information Integration Initiative (MI$^2$I) project 
of the Support Program for Starting Up Innovation Hub from Japan Science and Technology Agency (JST). 
R.M. is grateful for financial supports from MEXT-KAKENHI (17H05478 and 16KK0097), 
from Toyota Motor Corporation, from I-O DATA Foundation, 
and from the Air Force Office of Scientific Research (AFOSR-AOARD/FA2386-17-1-4049).
R.M. and K.H. are also grateful to financial supports from MEXT-FLAGSHIP2020 (hp170269, hp170220).

%%%%%%%%%%%%%%%%%%%%%%%%%%%%%%%
\section{Author Contributions}
%%%%%%%%%%%%%%%%%%%%%%%%%%%%%%%
Data curation, Tom Ichibha, Ornin Srihakulung, Guo Chao, and Ryo Maezono; Investigation, Tom Ichibha, Ornin Srihakulung and Guo Chao; Supervision, Luckhana Lawtrakul, Kenta Hongo and Ryo Maezono; Validation, Tom Ichibha, Ornin Srihakulung, Guo Chao and Adie Tri Hanindriyo; Visualization, Ornin Srihakulung; Writing - original draft, Tom Ichibha; Writing - review \& editing, Tom Ichibha, Adie Tri Hanindriyo, Luckhana Lawtrakul, Kenta Hongo and Ryo Maezono

%%%%%%%%%%%%%%%%%%%%%%%%%%%%%%%
\bibliographystyle{apsrev4-1}
\bibliography{references}
\end{document}